\documentclass[final,5p,times,twocolumn]{elsarticle}

\usepackage{graphicx}
\usepackage{amsmath,amssymb,amsfonts,amsbsy}
\usepackage[active]{srcltx}

\newcommand{\mathe}{\ensuremath{{\rm e}}}

\newcommand{\Mn}[1]{\ensuremath{{}^{#1}}{\rm Mn}}
\newcommand{\Eu}[1]{\ensuremath{{}^{#1}}{\rm Eu}}
\newcommand{\Au}[1]{\ensuremath{{}^{#1}}{\rm Au}}
\newcommand{\Si}[1]{\ensuremath{{}^{#1}}{\rm Si}}
\newcommand{\Ra}[1]{\ensuremath{{}^{#1}}{\rm Ra}}
\newcommand{\Pb}[1]{\ensuremath{{}^{#1}}{\rm Pb}}
\newcommand{\U}[1]{\ensuremath{{}^{#1}}{\rm U}}

\journal{NIM-A}

\begin{document}

\begin{frontmatter}



\title{Study of the dependence of $^{198}$Au half-life on source geometry}


\author[NIST]{R. M. Lindstrom}
\address[NIST]{Analytical Chemistry Division, National Institute of Standards and Technology, Gaithersburg, MD, 20899, USA}

\author[purdue]{E. Fischbach}

\address[purdue]{Physics Department, Purdue University, 525 Northwestern Avenue, West Lafayette, IN, 47907, USA}

\author[purdue]{J. B. Buncher}

\author[UT,ORNL]{G. L. Greene}
\address[UT]{University of Tennessee, Knoxville, TN, 37996, USA}
\address[ORNL]{Oak Ridge National Laboratory, Oak Ridge, TN, 37830, USA}

\author[NUCL,purdue]{J. H. Jenkins\corref{ca}}
\cortext[ca]{Corresponding Author}
\ead{jere@purdue.edu}
\address[NUCL]{School of Nuclear Engineering, Purdue University, 400 Central Dr, West Lafayette, IN, 47907, USA}

\author[wabash,purdue]{D. E. Krause}
\address[wabash]{Department of Physics, Wabash College, Crawfordsville, IN, 47933, USA}

\author[purdue]{J. J. Mattes}

\author[UT]{A. Yue}

\begin{abstract}
We report the results of an experiment to determine whether the half-life of \Au{198} depends on the shape of the source.  This study was motivated by recent suggestions that nuclear decay rates may be affected by solar activity, perhaps arising from solar neutrinos.  If this were the case then the $\beta$-decay rates, or half-lives, of a thin foil sample and a spherical sample of gold of the same mass and activity could be different.  We find for \Au{198}, $(T_{1/2})_{\rm foil}/(T_{1/2})_{\rm sphere} = 0.999 \pm 0.002$, where $T_{1/2}$ is the mean half-life.  The maximum neutrino flux at the sample in our experiments was several times greater than the flux of solar neutrinos at the surface of the Earth.  We show that this increase in flux leads to a significant improvement in the limits that can be inferred on a possible solar contribution to nuclear decays.
\end{abstract}

\end{frontmatter}


Extensive research has shown that the decay rate of a radioactive nucleus is independent of its environment, except in those instances involving electron capture, internal conversion, or high external magnetic fields
\cite{Hahn:1976}.  Therefore, it is surprising that a recent series of papers \cite{Jenkins:2008,Jenkins:2008a,Jenkins:2009,Jenkins:2009a,Fischbach:2009a} have raised the question of whether nuclear decay rates as observed in terrestrial laboratories are being influenced by solar activity.   These authors point to correlated seasonal variations
of the decay rates of \Si{32} at Brookhaven National Laboratory (BNL) \cite{Alburger:1986}, and of \Eu{152}, \Eu{154}, and \Ra{226} at the Physikalisch-Technische-Bundesanstalt (PTB) in Germany \cite{Schrader:1998,Schrader:2008}.  In addition, they found a statistically significant drop in the counting rate of \Mn{54} coincident in time with the solar flare of 2006 December 13 \cite{Jenkins:2008,Jenkins:2009,Fischbach:2009a}.

Although it may be the case that the observed fluctuations in the BNL/PTB data reflect some poorly understood instrumental effect, such as a seasonal dependence of detector efficiencies on external temperature, in the absence of additional evidence it remains an experimental question as to whether or not solar activity is in fact affecting the intrinsic decay rates. If this were the case then we might expect a range of responses from different nuclides arising from the same details of nuclear structure which account for the wide range of observed half-life ($T_{1/2}$) values. Given the unknown origin of the effect, it is thus possible that analyses of other data motivated by Refs.~\cite{Jenkins:2008,Jenkins:2008a,Jenkins:2009,Jenkins:2009a,Fischbach:2009a} have either not seen \cite{Cooper:2009,Norman:2009} or observed \cite{Ellis:1990} time-varying decay rates consistent with this work.

To account for the data in Refs.~\cite{Jenkins:2008,Jenkins:2008a,Jenkins:2009,Jenkins:2009a,Fischbach:2009a}, it was proposed that unstable terrestrial nuclei are being influenced by an (as yet unknown) interaction with solar neutrinos \cite{Fischbach:2009a}.   
A direct test of this hypothesis would be to measure decay rates of nuclei placed in a high flux of neutrinos such as produced by a nuclear reactor.   Preliminary work for such an experiment has already been carried out at the Pennsylvania State University Breazeale Nuclear Reactor.  Although initial results from this experiment were inconclusive due to unexpectedly large gamma backgrounds, the feasibility of such an experiment was established.  (It is interesting to note that in 1922 Hevesy \cite{Hevesy:1922} searched for modifications of the decay rates of \U{238}  and \Pb{210} samples that were irradiated with radium; no effect was observed.)   

The subject of this paper is an alternative approach to search for such a new neutrino interaction.  A consequence of the suggested interaction is that antineutrinos emitted by a sample of $\beta$-decaying nuclei should affect the decay rates of the remaining nuclei in a sample.  Interestingly, the flux of neutrinos that can be achieved in such an experiment (see below), is actually \textit{greater} that the estimated flux in the previously described reactor experiment.  In such a case the decay parameter $\lambda = \ln2/T_{1/2}$ would become time-dependent, reflecting the decrease in the antineutrino flux accompanying the decrease in the number of decaying nuclei.  For samples of similar total activity, this ``self-induced decay'' (SID) effect would depend on the shape of the sample, since antineutrinos produced in different geometries (for example a thin foil a sphere) would have different probabilities of influencing the surviving nuclei before escaping from the source. The objective of the experiment described below is to search for such a SID effect, and in the process address the more generic question of whether the half-life of source depends on its shape. Although this is obviously a fundamental question in $T_{1/2}$ metrology, it appears not to have been specifically addressed previously in the literature via a dedicated experiment.  Such an experiment could also set constraints on other new short-range interactions between nuclei that would affect the decay rates of neighboring nuclei.

We begin by developing the phenomenology used to characterize these experiments.  According the usual radioactivity decay law, the activity of a sample containing $N(t)$ unstable nuclei of the same type is $-dN/dt  \equiv -\dot{N}(t) = \lambda N(t)$, where $\lambda$ is a constant decay rate corresponding to these nuclei.  To parametrize a possible SID effect, let us assume that the decay parameter $\lambda$ for an individual nucleus will depend linearly on the local antineutrino density $\rho_{\nu}(\vec{r},t)$.  A more general dependence could be handled in a similar fashion.  We can write the decay rate of a nucleus at position $\vec{r}$ and time $t$ as $\lambda(\vec{r},t) = \lambda_0 + \lambda_1(\vec{r},t)$, where $\lambda_0$ represents the intrinsic contribution to the $\beta$-decay rate arising from the conventional weak interaction, along with a possible constant background arising from new interactions.  The perturbation $\lambda_1(\vec{r},t)$ is proportional to the local neutrino density, $\lambda_{1}(\vec{r},t) = \eta\rho_{\nu}(\vec{r},t)$, where the proportionality constant  $\eta$  depends on the strength of the interaction.

In the presence of the SID effect, the activity of a radioactive sample no longer obeys the simple decay law; since each nucleus experiences a different local neutrino density, the nuclei in the sample do not have the same decay rate.  However, the extensive experimental work mentioned above shows that if the SID effect is real, it must be very small to have gone undetected, which implies $|\lambda_1(\vec{r},t)| \ll \lambda_0$.  This allows one to define a sample-averaged decay rate $\overline{\lambda}(t) \equiv \lambda_{0} + \overline{\lambda}_{1}(t)$, where 
\begin{equation}
\overline{\lambda}_{1}(t) \equiv \frac{1}{N(t)} \int d^{3}r\, \rho(\vec{r},t)\lambda_{1}(\vec{r},t).
\end{equation}
Here $\rho(\vec{r},t)$ is the number density of undecayed nuclei and the integral is over the volume of the sample.  To lowest order in $\eta$, the neutrino density is proportional to $N(t)$ so $\overline{\lambda}_{1}(t)$ is proportional to $N(t)$.  We can thus write $\overline{\lambda}_1(t) \cong \xi\lambda_{0} N(t)/N_{0}$, where $N(t = 0) = N_{0}$, and  the dimensionless parameter $\xi$ depends on the geometry of the sample.  Notice that $\xi = \overline{\lambda}_{1}(0)/\lambda_{0}$ represents the fractional change in the decay rate at $t = 0$ due to the SID effect.  It follows that with $\overline{\lambda}_1(t)$ the decay law assumes the approximate form:
\begin{eqnarray}
 -\frac{dN(t)}{dt} = \lambda_0 N(t)\left[ 1 + \xi \frac{N(t)}{N_0} \right]. 
 \label{eqn:sid}
\end{eqnarray}
Eq.~\eqref{eqn:sid} can be solved exactly by assuming that $N(t) = N_0 \exp(-\lambda_0 t)F(t)$, in which case $F(t)$ is a solution of the equation $dF/F^2 = -\xi \exp(-x) dx$, where $x = \lambda_0 t$.  After solving for $F(x)$, we find
\begin{eqnarray}
 N(x) = \frac{N_0 \mathe^{-x}}{1 + \xi(1 - \mathe^{-x})}. 
 \label{eqn:nofx}
\end{eqnarray}
We note in passing that Eqs.~\eqref{eqn:sid} and \eqref{eqn:nofx} also describe the nonlinear growth of a biological population if appropriate signs for $\lambda_0$ and $\xi$ are incorporated \cite{Lotka:1956}.

While $\xi$ is a parameter that depends on the geometry of the sample under study,  we can estimate $\xi$ for a homogeneous spherical body of radius $R$.  One can show that for isotropic emission, the local density of neutrinos at a distance $r$ from the center is
\begin{eqnarray}
 \rho_{\nu}(r,t) = \frac{\lambda_0 \bar{\rho}}{c}\left[ \frac{R}{2} + \frac{R}{4}\left( \frac{R}{r} - \frac{r}{R}\right)\ln\left(\frac{R+r}{R-r} \right)\right],
\end{eqnarray}
where $\bar{\rho}(t) = 3 N(t)/(4 \pi R^3)$ is the average density of undecayed nuclei in the sample, and $c$ is the speed of light.  (To lowest order in $\eta$, we can neglect the spatial variation of $\overline{\rho}$ induced by the SID effect.)  Using this result, we find that the decay rate for the spherical sample is given by Eq.~\eqref{eqn:sid} with $\xi \rightarrow \xi_s$, where
\begin{eqnarray}
 \xi_s = \frac{\eta}{cR^2}\frac{9}{16 \pi}N_0.
 \label{eqn:betaeta}
\end{eqnarray}
For the foil, the corresponding quantity $\xi_f$ can be obtained by numerical integration.

\begin{figure}
 \includegraphics[height=54mm]{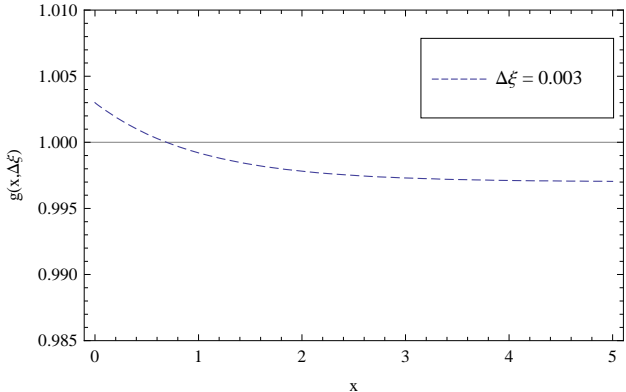}
 \caption{Plot of $g(x,\Delta\xi)$ from Eq.~\eqref{eqn:gofxb} with $\Delta\xi = 0.003$ and $N_{0s}/N_{0f} = 1$.}
 \label{fig:ratioalpha}
\end{figure}

We have carried out an experimental search for the SID effect by comparing the activities of \Au{198} ($T_{1/2} = 2.7$~d) as determined from a thin Au foil and from a sphere of the same mass and similar activity.  In this experiment the expression of interest is the ratio of count-rates for the sphere and foil as a function of $t = x/\lambda_0$.  Using Eq.~\eqref{eqn:nofx}, this is given (to first order in $\xi$) by
\begin{eqnarray}
 g(x, \Delta\xi) = \frac{\dot{N}_{\rm sphere}}{\dot{N}_{\rm foil}} \simeq \frac{N_{0s}}{N_{0f}}\left(1 - \Delta\xi + 2\Delta\xi e^{-x}\right), 
 \label{eqn:gofxb}
\end{eqnarray}
where $N_{0s}$ and $N_{0f}$ denote the number of \Au{198} atoms initially present in the sphere and foil, respectively, and $\Delta\xi = \xi_{s} - \xi_{f}$ is the difference in $\xi$ for the sphere and foil.   A plot of $g(x,\Delta\xi = 0.003)$ is shown for illustration in Fig.~\ref{fig:ratioalpha}, where the value of $\Delta\xi$ has been suggested by the data of Refs.~\cite{Jenkins:2008,Jenkins:2008a,Jenkins:2009,Jenkins:2009a,Fischbach:2009a}.  We see from Eq.~\eqref{eqn:gofxb} that for $N_{0s}/N_{0f} = 1$, $g(x=0, \Delta\xi) = 1 + \Delta\xi$, while $g(x\rightarrow\infty,\xi) = 1 - \Delta\xi$.  These results can be understood by noting that for $\Delta\xi > 0$ and $N_{0s}/N_{0f} = 1$ the increased decay rate arising from the SID effect causes a relative depletion of the \Au{198} population in the sphere initially compared to the foil, such that eventually a crossover in decay rates is achieved at $ t \cong T_{1/2}$.  Thereafter the relatively larger surviving population of decaying atoms in the foil produces a larger count rate in the foil which persists as $t \rightarrow \infty$.

As noted above, this experiment can be viewed as a generic test of whether nuclear decay rates depend on the shape of the sample, irrespective of the underlying mechanism.  However, if we are specifically interested in whether a shape-dependent SID effect could arise from antineutrinos emitted in $\beta$-decay then we must account for possible contributions to $\lambda_0$ from the solar neutrino flux $\mathcal{F}_{\odot} \cong 6 \times 10^{10}$~cm${}^{-2}$s${}^{-1}$.  
It is evidently desirable to start with samples having the largest possible internal antineutrino flux in order to achieve the largest effect.  Although the initial antineutrino flux in a spherical sample varies as a function of the distance $r$ from the center, a useful reference value is the flux at the surface $r = R$ which is given by
\begin{eqnarray}
 \mathcal{F}(R) = \frac{|dN/dt|}{4 \pi R^2} = \frac{\epsilon \lambda \rho R}{3} = 5.9 \times 10^{16} \,{\rm cm}^{-3}{\rm s}^{-1} \epsilon R,  \label{eqn:surfaceflux}
\end{eqnarray}
where $\lambda = 3.0 \times 10^{-6}$~s${}^{-1}$ is the \Au{198} decay constant, $\rho = 5.88 \times 10^{22}$~cm${}^{-3}$ is the Au atom density, and $\epsilon$ = $N_0(\Au{198})/N_0({\rm total})$.  It follows from Eq.~\eqref{eqn:surfaceflux} that $\mathcal{F}(R) \cong \mathcal{F}_{\odot}$ can be achieved for a sphere with $R = 1$~mm if $\epsilon \cong 1 \times 10^{-5}$, and these values guided the design of our experiment.  It should be noted, however, that since the effect of neutrinos may depend on their energies as well as their flux, choosing parameters such that $\mathcal{F}(R) > \mathcal{F}_{\odot}$ does not necessarily guarantee that the SID neutrinos will give the dominant contribution. We note in passing that since the specific activity achieved in this experiment exceeds that from previous measurements of the \Au{198} half-life, the conclusions drawn from earlier experiments are not directly relevant in the present context.

Two separate sphere:foil comparisons were performed. Squares were cut from pure Au foil (25~$\mu$m thick by 1.4~mm square), each of mass 1~mg.  Spheres were formed from the squares by melting in a borax flux. Each specimen was irradiated in the RT2 facility \cite{Lindstrom:2008} of the NIST Center for Neutron Research for 12 hours at a thermal neutron fluence rate of $3.1 \times 10^{17}$~m${}^{-2}$s${}^{-1}$ to produce an activity of 1.1~GBq (30~mCi) at the end of irradiation. Under these conditions the contribution to the total activity of the double-capture product \Au{199} ($T_{1/2}=3.1$~d) is calculated to be 1.5\%.  For the spherical 1~mg sample, $\mathcal{F}(R) \cong 2.7 \mathcal{F}_{\odot}$ initially, decreasing to $\mathcal{F}_{\odot}$ at $t \cong T_{1/2}$.

The radioactive decay of each sample was followed by repeated counting, using a Ge gamma spectrometry system that had been used previously to measure the half-life of \Au{198} \cite{Lindstrom:2005}.  Sources were held 40~cm from the detector end cap, with an absorber of 2.5~cm of Pb between the source and the detector. Spectra were collected every 2~hr (live time), with 65 to as many as 200 spectra in a data set. The initial counting rate was such that between $1 \times 10^6$ and $6 \times 10^6$ net counts were collected in the main gamma peak ($411.8$~keV) in each spectrum.

Net areas of the three \Au{198} gamma-ray peaks (411.8, 675.9, and 1087.7~keV) were integrated with a fixed-boundary baseline subtraction routine \cite{Lindstrom:1994}.  The same peak and baseline channels were used for all four data sets. Because of the thick absorber, the weaker higher energy lines were enhanced relative to the dominant 412~keV line and the low-energy gammas of \Au{199} were completely undetectable.

The decay data were fitted to an exponential decay law, correcting each datum for a linear combination of decay during the counting interval, extending dead time (pulse pileup in the analog circuitry) and non-extending dead time (analog-digital converter (ADC)) \cite{Fleming:1987,Lindstrom:1995}. The fit employed the Solver function in Microsoft Excel \cite{Fylstra:1998}, a nonlinear reduced gradient method for $\chi^2$ minimization to determine the decay constant $\lambda$, the initial activity, and the pileup parameter $\alpha$. The data points were weighted as the inverse square of the Poisson counting precision.  Individual fits were performed for each of three gamma rays in each of four samples (two foils and two spheres). To accentuate any effect of the highest specific activity, the twelve fits were repeated for a subset of the first thirty spectra, covering approximately the first half-life.

\begin{table*}
\centering
\caption{Half-lives (in hours)  determined for each gamma-ray in each sample, using all $n$ data points. The symbol $\pm$ here and throughout this work denotes the standard ($1 s$) uncertainty.}
\label{tab:table1}
\begin{tabular*}{\textwidth}{@{\extracolsep{\fill}}ccccc} \hline \hline 
Energy (keV) & Foil F1 ($n = 99$) & Sphere S1 ($n = 65 $)& Foil F2 ($n = 146$)& Sphere S2 ($n = 197$) \\ \hline
412  & $64.517 \pm 0.019$ & $64.709 \pm 0.046$ & $64.573 \pm 0.016$ & $64.589 \pm 0.009$ \\
676  & $64.504 \pm 0.061$ & $64.895 \pm 0.150$ & $64.554 \pm 0.040$ & $64.562 \pm 0.024$ \\
1088 & $64.705 \pm 0.095$ & $64.790 \pm 0.220$ & $64.716 \pm 0.057$ & $64.584 \pm 0.033$ \\
\hline
$T_{1/2}^{\rm Foil}/T_{1/2}^{\rm Sphere}$ & \multicolumn{2}{c}{$0.9968 \pm 0.0014$}  & \multicolumn{2}{c}{$0.9999 \pm 0.0007$} \\
\hline \hline
\end{tabular*}
\end{table*}
\begin{table*}
\centering
\caption{Half-lives (in hours)  determined from the initial $n = 30$ spectra (first 63 hours).}
\label{tab:table2}
\begin{tabular*}{\textwidth}{@{\extracolsep{\fill}}ccccc} \hline \hline 
Energy (keV) & Foil F1 & Sphere S1 & Foil F2 & Sphere S2  \\ \hline
412  & $64.92 \pm 0.24$ & $64.38 \pm 0.24$ & $63.69 \pm 0.39$ & $64.82 \pm 0.44$ \\
676  & $63.76 \pm 0.73$ & $64.12 \pm 0.73$ & $63.77 \pm 1.05$ & $66.82 \pm 1.24$ \\
1088 & $64.32 \pm 1.08$ & $65.08 \pm 1.20$ & $65.12 \pm 1.49$ & $66.84 \pm 1.73$ \\
\hline
$T_{1/2}^{\rm Foil}/T_{1/2}^{\rm Sphere}$ & \multicolumn{2}{c}{$1.006 \pm 0.008$}  & \multicolumn{2}{c}{$0.979 \pm 0.008$} \\
\hline \hline
\end{tabular*}
\end{table*}

To minimize rate-related instrumental effects, the data were first examined in a model-free way, simply comparing peak areas of the sphere to the foil in pairs of spectra measured at nearly the same counting rate (but slightly different decay times). A small temporal trend in the gamma peak ratios for the first experiment was not significantly different from that of the associated pulser peak. In the second experiment there was no detectable difference in $149$ sphere/foil ratios at $412$~keV over nearly $12$~half-lives.

More fundamentally, half-lives were determined by applying the model described in Refs.\cite{Fleming:1987,Lindstrom:1995} to each data set: two experiments, two samples each (sphere vs.~foil), and three gamma rays per sample. The uncertainty ($1 s$) in the half-life was assigned as the increment that increased the weighted sum of squares of normalized residuals by one unit \cite{Rogers:1975}.  The model described the data very well: the median $\chi^2$ per degree of freedom was $1.1$ (range $0.9$ to $2.6$). The results are given in Table~\ref{tab:table1}.

To search for effects at the highest specific activity, the decay curves were re-fitted using a subset of the data encompassing the first half-life.  Because of the shorter decay duration, the uncertainty of the half-life was greater, as shown in Table~\ref{tab:table2}.  Half-lives measured in this work are in agreement with those previously determined in this laboratory \cite{Lindstrom:2005}.

The measurements in Tables \ref{tab:table1} and \ref{tab:table2} are dominated by the strong $412$~keV peak.  For the full data sets, the weighted mean foil/sphere half-life ratio is $0.9968(14)$ for experiment 1 and $0.9999(7)$ for experiment 2, where the uncertainty is the standard deviation of the mean.  Assuming a normal distribution, the probabilities that the ratio does not differ significantly from unity are $P = 0.02$ and $P = 0.90$, respectively.  The second experiment, with nearly twice the number of data points, shows less difference between the two shapes.


Our results can be used to estimate $\Delta \xi$ by noting from Eq.~\eqref{eqn:gofxb} that
\begin{eqnarray}
 \frac{g(x=0,\Delta \xi)}{g(x\rightarrow \infty,\Delta \xi)} = \frac{(\dot{N}_{\rm sphere}/\dot{N}_{\rm foil})_{t=0}}{(\dot{N}_{\rm sphere}/\dot{N}_{\rm foil})_{t \rightarrow \infty}} \cong 1 + 2 \Delta \xi. \label{eqn:gratio}
\end{eqnarray}
For each of experiments 1 and 2 the $t = 0 \, (t \rightarrow \infty)$ ratio was obtained by averaging the 411.8~keV data over five 2-hour counts at the beginning (end) of the experiment.  We find for the ratio in Eq.~\eqref{eqn:gratio} the value $0.9964(7)$ from experiment 1, and $0.9992(16)$ from experiment 2; the weighted mean of both experiments gives $0.9969(10)$, where the $1\sigma$ uncertainties are obtained from propagated Poisson statistics.  Inserting these results into Eq.~\eqref{eqn:gratio} yields $\Delta \xi_1 = -1.8(4) \times 10^{-3}$, $\Delta \xi_2 = -4(8) \times 10^{-4}$, and $\Delta \xi_m = -1.55(50) \times 10^{-3}$, obtained from experiments 1, 2, and their mean, respectively.  For purposes of comparison, the peak-to-trough variation in the BNL Si/Cl ratio of count rates, $\Delta r$ \cite{Jenkins:2009a} is $\sim 3\times 10^{-3}$, corresponding to a change $\Delta f$ in solar flux of $6.683 \times 10^{-2}$ from perihelion to aphelion.  The ratio $\Delta r / \Delta f \approx 4 \times 10^{-2}$ provides a measure of the improved sensitivity of the current experiment when compared to the above values of $\Delta \xi$.

Although the preceding comparisons of the $t = 0$ and $t \rightarrow \infty$ count-rates exhibit small deviations from the expected null results, the overall half-lives measured in this work are in agreement with those previously measured in this laboratory \cite{Lindstrom:2005}.  From the full data sets the mean foil/sphere half-life ratio is $0.999 \pm 0.002$, and for the initial decay period the ratio is $1.00 \pm 0.02$.  These results are consistent with the deviations previously noted in $\Delta \xi_1$ and $\Delta \xi_m$:  returning to Fig.~\ref{fig:ratioalpha} we note that there is a degree of cancellation that arises when $T_{1/2}$ for the spherical samples is obtained by combining data for $t < T_{1/2}$ with data for $t > T_{1/2}$.  Separating the data from these two time intervals is thus always desirable, but comes at a cost in statistics, as we have noted previously.

In summary, the present experiment is the first direct precision test of whether the decay rate of a radioactive source depends on its shape.  Our results in Table \ref{tab:table1} indicate a $\sim \! 2.3 \sigma$ deviation of the foil/sphere ratio in experiment 1 from unity.  From Table \ref{tab:table2}, based on the initial 30 spectra, the foil/sphere ratio for experiment 2 deviates from unity by $\sim \! 2.6 \sigma$.  These results thus leave open the possibility that the half-life of a radioactive nuclide could in fact depend on its shape (due to the internal flux of neutrinos, photons, or electrons), and hence suggests that additional experiments are necessary.  We are thus planning to repeat this experiment in an attempt to reduce our uncertainties by starting with foil and sphere samples with greater activity.  If a significant effect attributable to neutrinos were seen in such an experiment, this would support the inference drawn from Ref.\cite{Jenkins:2008,Jenkins:2008a,Jenkins:2009,Jenkins:2009a,Fischbach:2009a} of a possible solar influence on nuclear decay rates.

The authors wish to thank R. Reifenberger for his assistance in the early stages of this collaboration, and A. Rukhin for statistical discussions.  The work of EF was supported by the U.S. Department of Energy under contract No. DE-AC02-76ER071428, and the work of GLG was supported in part by the U.S. Department of Energy under grant DE-FG02-03ER41258.  Disclaimer: The identification of any commercial product or trade name does not apply endorsement or recommendation by the authors or their institutions.  Contributions of the National Institute of Standards and Technology are not subject to U.S. copyright.


\begin{thebibliography}{10}
\expandafter\ifx\csname url\endcsname\relax
  \def\url#1{\texttt{#1}}\fi
\expandafter\ifx\csname urlprefix\endcsname\relax\def\urlprefix{URL }\fi
\expandafter\ifx\csname href\endcsname\relax
  \def\href#1#2{#2} \def\path#1{#1}\fi


\bibitem{Hahn:1976}
H.-P. Hahn, H.-J. Born, J.~Kim, Survey on the rate of perturbation of nuclear
  decay, Radiochimica Acta 23 (1976) 23--37.


\bibitem{Jenkins:2008}
J.~Jenkins, E.~Fischbach, {Perturbation of Nuclear Decay Rates During the Solar
  Flare of 13 December 2006}, arXiv:0808.3156v1 [astro-ph].


\bibitem{Jenkins:2008a}
J.~Jenkins~et al., {Evidence for Correlations Between Nuclear Decay Rates and
  Earth-Sun Distance}, arXiv:0808.3283v1 [astro-ph].


\bibitem{Jenkins:2009}
J.~Jenkins, E.~Fischbach, {Perturbation of Nuclear Decay Rates During the Solar
  Flare of 13 December 2006}, Astropart. Phys. 31 (2009) 407.


\bibitem{Jenkins:2009a}
J.~Jenkins, E.~Fischbach, J.~Buncher, J.~Gruenwald, D.~Krause, J.~Mattes,
  {Evidence for Correlations Between Nuclear Decay Rates and Earth-Sun
  Distance}, Astropart. Phys. 32 (2009) 42--46.


\bibitem{Fischbach:2009a}
E.~Fischbach, J.~Buncher, J.~Gruenwald, J.~Jenkins, D.~Krause, J.~Mattes,
  J.~Newport, {Time-Dependent Nuclear Decay Parameters: New Evidence for New
  Forces?}, Space Sci. Rev. 145 (2009) 285--335.


\bibitem{Alburger:1986}
D.~E. Alburger, G.~Harbottle, E.~F. Norton, Half-life of $^{32}${S}i, Earth and
  Planetary Sci. Lett. 78 (1986) 168--176.


\bibitem{Schrader:1998}
H.~Siegert, H.~Schrader, U.~Sch\"{o}tzig, {Half-life Measurements of Europium
  Radionuclides and the Long-term Stability of Detectors}, Appl. Radiat. Isot.
  49~(9-11) (1998) 1397.


\bibitem{Schrader:2008}
H.~Schrader, private communication (2008).


\bibitem{Cooper:2009}
P.~S. Cooper, {Searching for modifications to the exponential radioactive decay
  law with the Cassini spacecraft}, Astropart. Phys. 31 (2009) 267--269.


\bibitem{Norman:2009}
E.~B. Norman, E.~Browne, H.~Shugart, T.~Joshi, R.~Firestone, {Evidence Against
  Correlations Between Nuclear Decay Rates and Earth-Sun Distance}, Astropart.
  Phys. 31 (2009) 135--137.


\bibitem{Ellis:1990}
K.~J. Ellis, The effective half-life of a broad beam ${}^{238}${P}u{B}e total
  body neutron irradiator, Phys. Med. Biol. 35~(8) (1990) 1079--1088.


\bibitem{Hevesy:1922}
G.~Hevesy, An attempt to influence the rate of radioactive disintegrations by
  use of penetrating radiation, Nature 110~(2754) (1922) 216.


\bibitem{Lotka:1956}
A.~J. Lotka, Elements of Mathematical Biology, Dover, New York, 1956, p.65.


\bibitem{Lindstrom:2008}
R.~Lindstrom, R.~Zeisler, E.~Mackey, P.~Liposky, R.~Popelka-Filcoff,
  R.~Williams, {Neutron irradiation in activation analysis: A new rabbit for
  the NBSR}, J. Radioanal. Nucl. Chem. 278 (2008) 665--669.


\bibitem{Lindstrom:2005}
R.~Lindstrom, M.~Blaauw, M.~Unterweger, {The Half-Lives of ${}^{24}$Na,
  ${}^{43}$K, and ${}^{198}$Au}, J. Radioanal. Nucl. Chem. 263 (2005) 311--313.


\bibitem{Lindstrom:1994}
R.~Lindstrom, {SUM and MEAN: Standard Programs for Activation Analysis}, Biol.
  Trace Elem. Res. 43--45 (1994) 597--603.


\bibitem{Fleming:1987}
R.~Fleming, R.~Lindstrom, {Precise Determination of Aluminum by Instrumental
  Neutron Activation}, J. Radioanal. Nucl. Chem. 113 (1987) 35--42.


\bibitem{Lindstrom:1995}
R.~Lindstrom, R.~Fleming, {Dead Time, Pileup, and Accurate Gamma-Ray
  Spectrometry}, Radioact. Radiochem. 6~(2) (1995) 20--27.


\bibitem{Fylstra:1998}
D.~Fylstra, L.~Lasdon, J.~Watson, A.~Warren, {Design and use of the Microsoft
  Excel Solver}, Interfaces 28~(5) (1998) 29--55.




\bibitem{Rogers:1975}
D.~Rogers, Analytic and graphical methods for assigning errors to parameters
  in non-linear least squares fitting, Nucl. Inst. Meth. 127 (1975) 253.


\end{thebibliography}

\end{document}